\documentclass{ewshm}
\usepackage{graphicx}

\title{Optimal Meal Schedule for a Local Nonprofit Using LLM-Aided Data Extraction}

\author[1]{Sergio Marin}
\author[1]{Nhu Nguyen}
\author[1]{Max (Bohong) Zheng}
\author[1]{Christina M. Weaver}

\affiliation[1]{Department of Mathematics \& Statistics, Franklin \& Marshall College, 
Lancaster, PA, USA 17604}

\begin{document}

\maketitle 

\begin{abstract}
We present a data-driven pipeline developed in collaboration with the Power Packs Project, a nonprofit addressing food insecurity in local communities. The system integrates data extraction from PDFs, large language models for ingredient standardization, and binary integer programming to generate a $15$-week recipe schedule that minimizes projected wholesale costs while meeting nutritional constraints. All $157$ recipes were mapped to a nutritional database and assigned estimated and predicted costs using historical invoice data and category-specific inflation adjustments. The model effectively handles real-world price volatility and is structured for easy updates as new recipes or cost data become available. Optimization results show that constraint-based selection yields nutritionally balanced and cost-efficient plans under uncertainty. To facilitate real-time decision-making, we deployed a searchable web platform that integrates analytical models into daily operations by enabling staff to explore recipes by ingredient, category, or through an optimized meal plan.

\end{abstract}


\section*{Introduction}

Food insecurity, commonly defined as the lack of reliable access to affordable and nutritious food, remains a persistent challenge that affects millions of households in the United States, with serious implications for child development and family well-being~\cite{hines2021}. In 2023 alone, an estimated 18 million US households experienced food insecurity~\cite{usda2025}. The Power Packs Project (PPP), a nonprofit based in Lancaster, Pennsylvania, addresses this problem by distributing weekly food boxes and recipe cards to families with school-aged children in Lancaster, Lebanon, and York counties~\cite{ppp-site}. In the academic year 2024-25, PPP averaged more than 14,000 meals per week. By pairing donated or low-cost ingredients with simple and nutritious recipes, PPP not only improves food security, but also encourages families to cook and eat together, a practice associated with improved nutritional and psychosocial outcomes for children~\cite{hammons2011,fiese2006,smith2013}. Families who pick up their weekly Power Pack report less food-related stress and greater food sufficiency compared to those who do not~\cite{ryan2023}.

Despite its impact, PPP operates with a small paid staff and a large volunteer base, making efficiency crucial. Over the past six years, PPP has produced more than 150 recipes, stored as PDFs, many of which are reused or adapted depending on available inventory. Through partnerships and donations, ingredients are often acquired at low or zero cost, but unpredictable availability and inflation make it difficult to anticipate the cost of any given recipe.

Optimization methods such as linear programming have been widely applied in institutional contexts, such as schools and long-term care facilities, to generate low-cost nutritionally adequate meal plans~\cite{dooren2018}. Many of these models adopt multi-objective frameworks to jointly address cost, environmental impact, food waste, and nutritional quality. For instance, Vici~\cite{vici2025} optimized Italian primary school menus with respect to nutrition, sustainability, and food waste, while Benvenuti and De Santis~\cite{benvenuti2020} developed a binary linear programming model to produce balanced and cost-effective diets in schools and nursing homes. Despite their strengths, these systems are typically designed for fixed meal services with stable data inputs, which limits their applicability in dynamic, resource-constrained environments such as community nonprofits.

In parallel, advances in natural language processing, particularly large language models (LLMs), have enabled robust extraction and standardization of unstructured text data, including ingredient strings. Applications of LLMs in the food domain are emerging, such as food name entity linking~\cite{feher2023}, but few tools exist that apply these models in a practical, nonprofit setting for real-time nutritional and cost analysis.

In this study, we present a data-focused pipeline designed in close collaboration with the PPP team. Our system digitizes and cleans their full archive of recipes, uses LLM-based semantic matching to align ingredient strings with a nutritional database, and links each recipe to a predicted wholesale cost using historical invoices and category-specific inflation adjustments. A binary integer programming model then selects an optimized 15-week recipe schedule that satisfies key nutritional thresholds while minimizing expected cost. All data are integrated into a searchable web platform that allows PPP staff to plan by ingredient, category, or optimized schedule.

\begin{figure}[ht]
    \centering
    \includegraphics[width=\textwidth]{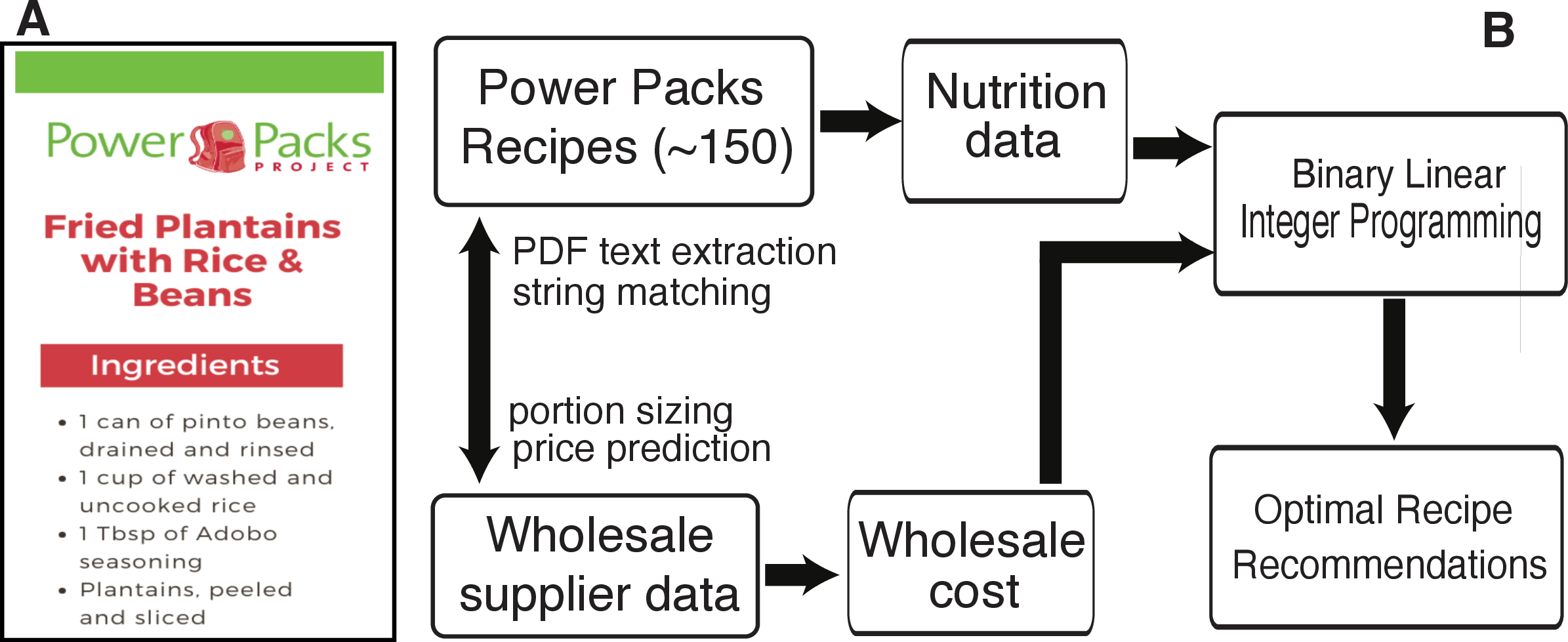} 
    \caption{\textbf{Recipe optimization workflow.} A: Excerpt from one recipe used for data extraction. B: Flowchart of our approach, starting from the provided recipes and invoices, to our optimal recipe recommendations. }
    \label{fig:flowcht}
\end{figure}

This work contributes a scalable, low-barrier system that blends modern NLP and optimization methods within a nonprofit's operational workflow. The pipeline is modular and extensible, supporting future use cases such as inventory-aware planning or dynamic cost tracking, and demonstrates how tailored data tools can support community-level food security with measurable impact. A visual summary of the full pipeline is shown in Figure~\ref{fig:flowcht}.

\newpage
\section{Methods}

All code was written in Python 3.12.5 and RStudio 4.4.2~\cite{python,rstudio}.

\subsection{Recipe Database Construction}

We compiled a comprehensive database of $157$ recipes distributed by PPP from the academic years $2019$--$2020$ through $2024$--$2025$, sourced from Microsoft Word and PDF documents.  Typically PDFs were read using standard Python tools. When necessary, optical character recognition (OCR) was used to extract text from non-selectable formats. The resulting dataset, in which each unique recipes is tagged with its distribution year and full ingredient list, was reviewed manually for errors.  Recipes were assigned to one of four categories based on the kind of protein it contained:  beef, poultry, seafood, or vegetarian.

\subsection{String-Matching for Nutritional Mapping}

To quantify nutritional data for each recipe, we first compared PPP recipe ingredient strings against more than $13,000$ entries in the freely-available My Food Data (MFD) nutritional database [link].  The matching process uses the \texttt{all-MiniLM-L6-v2} model from the 
\newline \texttt{SentenceTransformers} library in Python~\cite{sentence-transformers}, which encodes text strings into unique representations within a $384$-dimensional vector space. The semantic similarity between two strings $A$ and $B$ was measured as

\begin{equation}
sim(A,B)=\cos \theta = \frac{\mathbf{A} \cdot \mathbf{B}}{\|\mathbf{A}\| \, \|\mathbf{B}\|},
\label{eq:cosine}
\end{equation}

\noindent where $\theta$ is the angle between $A$ and $B$ in this embedded vector space.

We implemented a two-stage string matching approach, first identifying the string from MFD with highest similarity to a given PPP ingredient.  If visual inspection confirmed this match was suitable, the MFD identification number was saved.  Otherwise, the four next-best matches between the ingredient and MFD strings were identified, and the best match of these five candidates was identified visually.  If none of the five candidates were suitable, an appropriate match was identified by a manual search of MFD.

After obtaining suitable mappings between PPP ingredients and MFD entries, we converted nutritional information from MFD (reported per $100$ g) to reflect per-person serving sizes from the PPP recipes. Since recipe measurements were given in household units such as cups, ounces, or cans, ingredient weights could vary widely.  Recipes with missing ingredient measurements were updated by hand, based on measurements in similar recipes.  Ingredient categories with similar weights were grouped together manually, with reasonable measurement-to-weight conversion factors identified accordingly.  Nutritional values per serving for each recipe were then obtained by summing those from each ingredient.  

Nutrients obtained included calories; protein, fiber, carbohydrates, fat, saturated fats (measured in g); calcium, iron, magnesium, vitamin C, zinc, cholesterol, sodium (measured in mg); and folate as well as vitamins C, B$_6$ and B$_{12}$ (measured in mcg).  These were selected through consultation with a dietitian.

\subsection{Recipe Price Prediction}

To assess the cost of ingredients, we constructed a second, product pricing database using historical vendor invoice PDFs from four sources:  the Central Pennsylvania Food Bank, the Restaurant Store, and two local produce wholesalers. We had one year of data (2024--25) from all sources, plus three more years from the Central PA Food Bank (2021--24).  We extracted and cleaned product-level pricing data, categorizing items by food group. 
Then each recipe ingredient was matched manually to one or more entries from the pricing database, enabling estimation of the price per unit (or per pound, as appropriate) over time. 

PPP sometimes (but not always) obtained recipe ingredients for free.  We used this to construct a basic model of $C_i$, the cost of each ingredient $i$ for the 2025--26 academic year:

\begin{equation}
C_i=P(B_i)\cdot\bar{p}_i\cdot(1+I_i),
\label{eq:price}
\end{equation}

\noindent where
$B_i$ is the probability of buying the ingredient (estimated as the number of times PPP purchased the item, divided by the number of times they ordered it), 
$\bar{p}_i$ is its average historical unit price, and 
$I_i$ is its price inflation factor (Table \ref{tab:inf}; \cite{inflation-beef/poultry/fruits/seaood/vegetables, inflation-pasta/rice/flour, inflation-BeansLegumes/Seasonings/Soups/Sausage, inflation-dairy, inflation-eggs}
). Summing the predicted, per-unit cost of all ingredients in a recipe yields $R_j$, the inflation-adjusted predicted cost for recipe $j$. 

\begin{table}[ht]
\caption{Price inflation factors for various food categories. }
\label{tab:inf}
    \centering
    \begin{tabular}{|c|c|c|c|c|c|c|}
        \hline
    Product	& Beef & Poultry & Seafood & Eggs & Legumes	& Dairy \\ \hline
Factor	& 6.8	& 2.3	& 3	& 40	& 2.7	& 1.7 \\ \hline \hline
Product	& Pasta/Rice	& Fruit	& Vegetables	& Baking & Spices & Soups	 \\ \hline
Factor	& -0.6	& 1.6	& -2.5	& 1.5	& 1.6	& 1.6 \\ \hline
    \end{tabular}
    \end{table}

\subsection{Web Platform for Recipe Access and Nutrition/Cost Exploration}

A user-friendly web application was developed [https://powerpacks.dreamhosters.com/] to enable PPP to search and explore the recipe database. Users can search recipes by category or ingredients and view standardized ingredient lists. 

\subsection{Optimization of Future Recipe Schedule}

Once the database of nutrition and price information was constructed for all $N=157$ recipes, we applied binary linear integer programming to identify $M$ recipes which minimized the total cost while maintaining a basic level of nutrition and including recipes from each of the four protein categories.  Here we used $M=15$, representing the number of recipes that cover the first half of PPP's upcoming academic year. For flexibility in the flavor profiles of selected recipes, we chose to enforce only three broad nutritional constraints:  that the average amount of protein and calcium among selected recipes were above certain minimal values ($\bar{P}_{min}=15$~g and $\bar{C}_{min}=433.33$~mg respectively), and that the average amount of fat was below a certain maximum ($\bar{C}_{min}=30$ g).  These protein and calcium bounds are equal to one-third the daily intake recommended by USDA for 14--18 year-old children \cite{diet-guide}.  The maximal fat bound equals half of 60 grams, which is an estimate of the maximal daily fat intake recommended for an 1800-calorie diet ($\leq 30$\% of daily calories from fat \cite{fat-calc}).

Mathematically, letting \( x_j \in \{0,1\} \) denote whether recipe \( j \) was selected, we formulated the problem as follows:
\[
\begin{aligned}
\min \quad & \sum_{j=1}^N R_j x_j \\
\text{such that} \quad 
& \sum_{j=1}^N Prot(j)\cdot x_j \geq \bar{P}_{min} \cdot M \\
& \sum_{j=1}^N Cal(j)\cdot x_j \geq \bar{C}_{min} \cdot M \\
& \sum_{j=1}^N Fat(j)\cdot x_j \leq \bar{F}_{max} \cdot M, \\
& \sum_{j=1}^N x_j = M \\
& \left \lfloor \frac{M}{4} \right \rfloor \leq \sum x_{j \in \text{Category}_k} \leq \left \lfloor \frac{M}{4} \right \rfloor +1 \quad \forall k \in \{\text{Beef, Poultry, Seafood, Vegetarian}\}.\\
\end{aligned}
\]
\newline Here, $R_j$ is the predicted cost of recipe $j$, while $Prot(j)$, $Cal(j)$ and $Fat(j)$ represent the amounts of protein, calcium, and total fat in recipe $j$ respectively.  The floor functions ensure that the number of recipes from each protein category is approximately equal.  The solution to this linear program provides a nutritionally adequate and cost-effective meal plan that meets the operational needs of the Power Packs team.

\section{Results}

\subsection{Ingredient Matching and Nutrition Conversion Results}

Construction of the recipe database extracted from four years' worth of PPP's PDF files identified a total of $866$ ingredients across $157$ recipes.  Python's \texttt{SentenceTransformers} function used semantic embeddings to map each PPP recipe ingredient and My Food Database (MFD) entry onto a $384$-dimensional vector space.  Then the similarity measure predicts which of the MFD entries were most similar semantically to the recipe ingredient. As an example, Figure $2A$ shows the black vector corresponding to the PPP recipe ingredient ``Milk" along with different colored vectors corresponding to four phrases from MFD.  The two kinds of milk which are typically in liquid form ---as one would assume ``Milk" is in a PPP recipe---point in very similar directions to the Milk vector.  The two other MFD entries are pointing in very different directions.  Among these four MFD entries, ``Goat Milk" has highest semantic similarity to ``Milk" (0.82), though manually we would identify ``Whole Milk" as a better match.

\begin{figure}[ht]
    \centering
    \includegraphics[width=\textwidth]{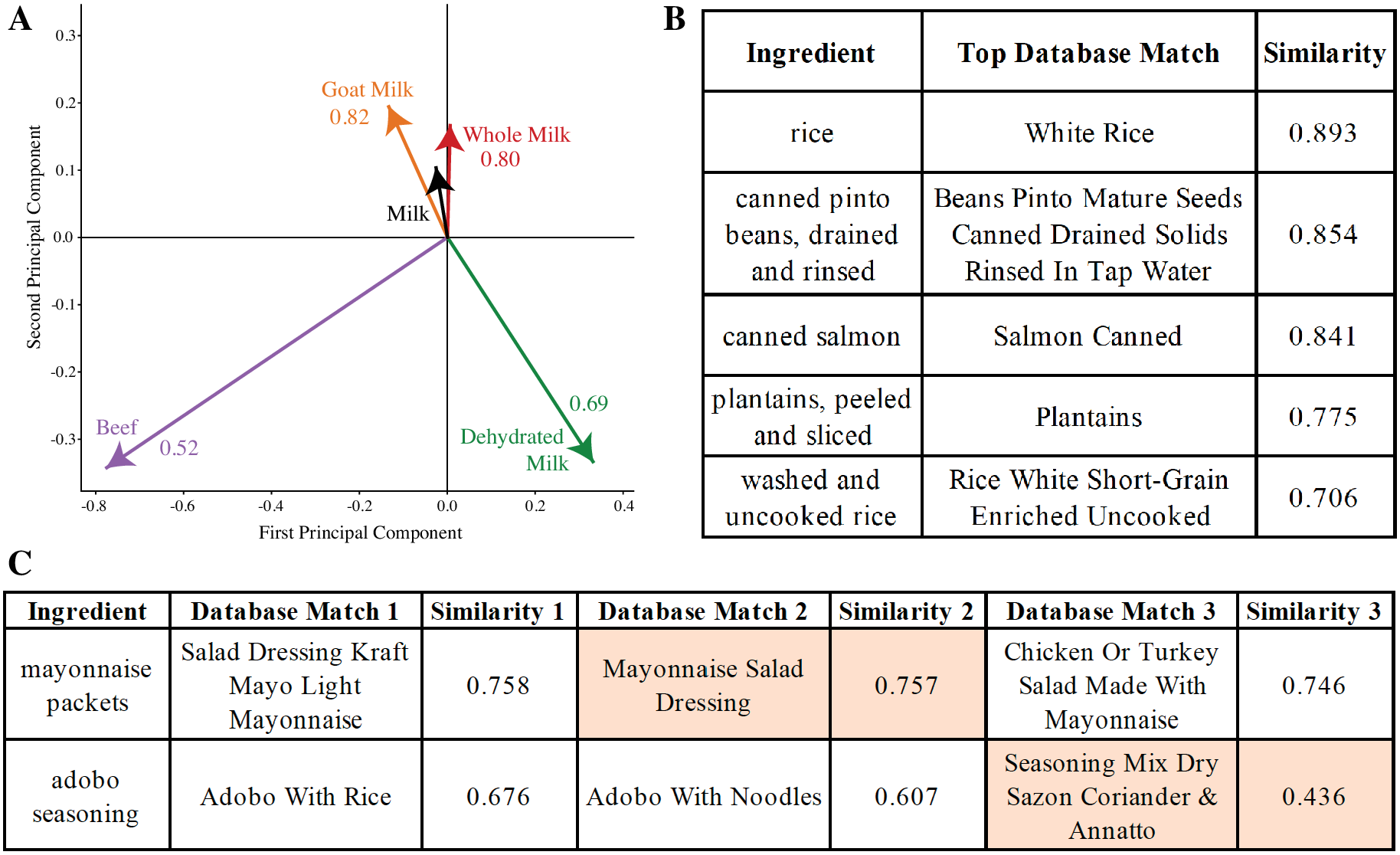} 
    \caption{\textbf{Matching strings of PPP recipe ingredients to the My Food Database.}  A: projection of five phrases onto the first two principal components of the SentenceTransformer vector space.  
    B-C:  Successful ingredient matches after stage one (B) or stage two (C).  Orange highlighting indicates manually identified best match.}
    \label{fig:strmtch}
\end{figure}

Our two-stage string-matching approach proved effective in aligning PPP recipe ingredients with MFD database entries. In total, $87\%$ of the ingredients received a valid match in the first stage. These matches demonstrated strong semantic alignment between the recipe terms and MFD entries, with high cosine similarity scores, as shown in Figure 2B.  Some of these semantically similar best matches had text strings that were nearly identical; in other cases, the actual strings were quite different.  

The remaining $13\%$ of ingredients required the second string-matching stage. While the similarity scores for the second through fifth candidates were generally lower than the top-ranked match, the lower-ranked options often provided better contextual alignment with the original recipe ingredient. In some cases, the best match had a similarity score below $0.5$. For example, in Figure 2C, the similarity score for ``Seasoning Mix Dry Sazon Coriander \& Annatto'' was $0.436$---the lowest among the top three candidates shown---but ultimately we decided it was the most appropriate match for our goal of estimating nutritional information of the original ``Adobo Seasoning" ingredient.  On the rare occasion that a recipe ingredient had no suitable match after stage two, we identified the best match among MDB entries manually.  

After finalizing MFD matches for all PPP ingredients, we proceeded to convert the MFD nutritional data as described above.  The result was a powerful and robust dataset that illustrated the nutritional diversity of PPP recipes (Fig. 3).  The scatterplot of calories versus fat revealed a positive but diffuse correlation. Recipes with higher values of both quantities were mostly found in the beef and poultry categories. However, a few vegetarian and seafood recipes also appeared in the upper-right quadrant as outliers, despite most of them clustering in the lower-left region with lower fat and calorie content. One of these outliers (solid green circle) was a Macaroni and Cheese recipe, selected by the optimization procedure below.  This variation highlights the trade-offs involved in selecting energy-dense versus leaner meals across different protein types, as part of a varied diet.

Examining calcium vs. protein (Fig. 3B) revealed that most beef, poultry, and seafood dishes fall to the right of the protein lower bound, while vegetarian recipes tend to lie on the left side, reflecting their generally lower protein content. Two clear outliers---one seafood and one vegetarian recipe---show unusually high calcium values.  Further inspection reveals that these two are Seafood Chowder and Macaroni and cheese recipes (Table 2). 


Boxplots of carbohydrates, cholesterol, and sodium, respectively, are shown in Figure 3C-E. For cholesterol and sodium, beef recipes exhibit the highest median values, while vegetarian meals show the lowest. This pattern is consistent with expectations based on ingredient types. However, the carbohydrate distribution reveals a different trend: vegetarian recipes tend to have the highest carbohydrate content on average. This reflects the reliance of vegetarian meals on grains, legumes, and starch-heavy components.

\begin{figure}[ht]
    \centering
    \includegraphics[width=\textwidth]{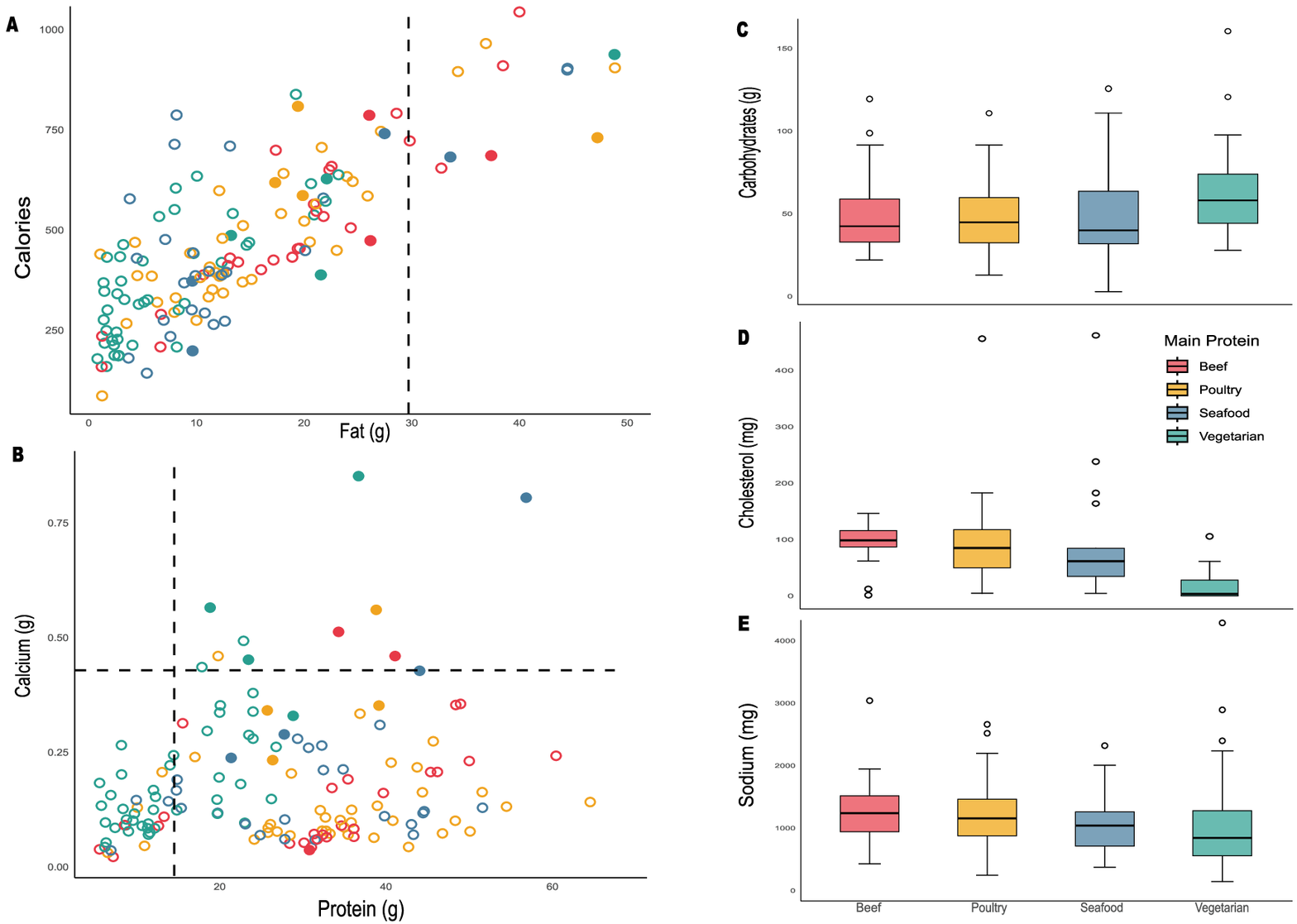} 
    \caption{\textbf{Nutritional Metrics of PPP Recipes.} A-B: Projections of PPP recipes onto Fat vs. Calories, and Protein vs. Calcium. Dashed lines indicate the bounds for average fat, calcium, and protein during optimization; filled circles indicate recipes included in the optimal schedule, C-E:  Distribution of other dietary elements, separated by protein category.}
    \label{fig:myimage}
\end{figure}

\subsection{Price Estimation and Prediction}

\begin{figure}[ht]
    \centering
    \includegraphics[width=\textwidth]{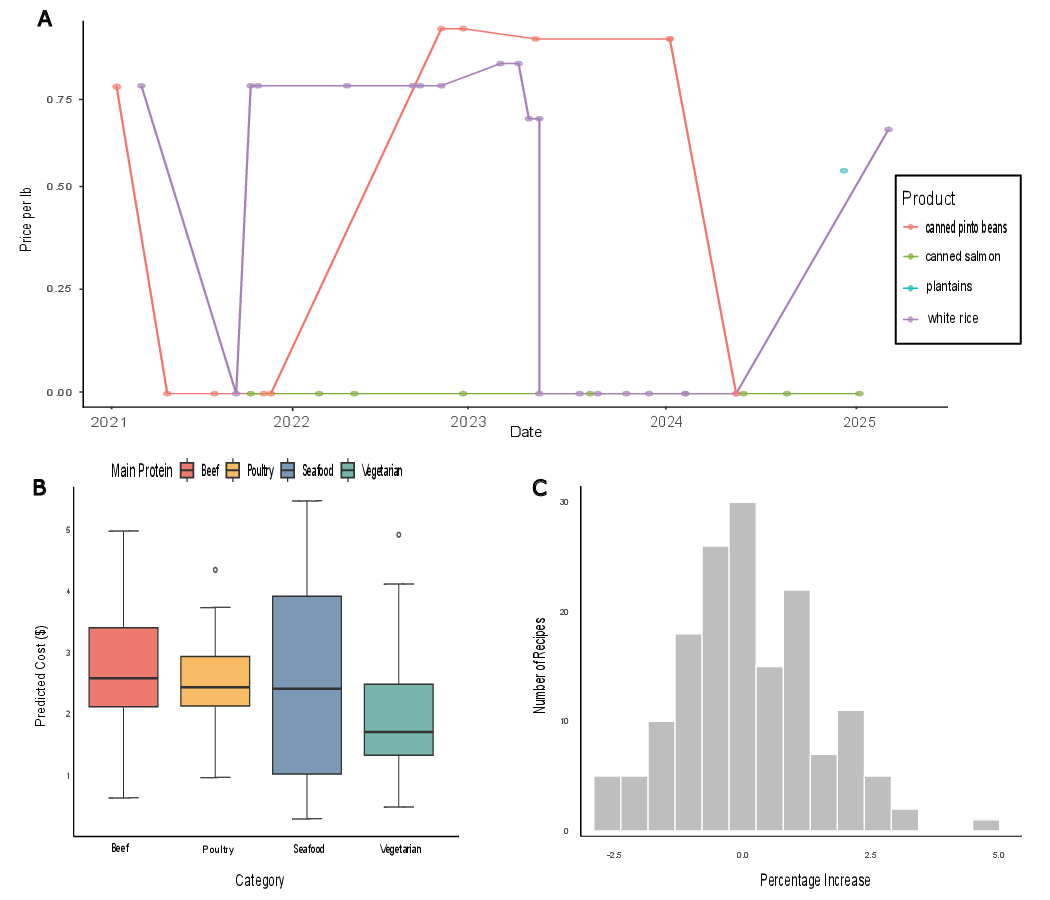} 
    \caption{\textbf{Price Estimation and Prediction.} \textbf{(A)} Historical prices per pound of selected ingredients. \textbf{(B)} Predicted recipe costs by protein category. \textbf{(C)} Histogram of percentage increase in predicted cost across recipes.}
    \label{fig:price_prediction}
\end{figure}

Historical invoice data revealed that wholesale prices of some recipe ingredients varied substantially over time (Fig. ~\ref{fig:price_prediction}A).  Some items, like canned salmon, were free consistently.  Others were used frequently but only free occasionally, including white rice and pinto beans; some (like plantains) were purchased but only rarely.  

Figure~\ref{fig:price_prediction}B compares the predicted cost of recipes across protein categories. Beef recipes exhibited the highest median cost and variability, closely followed by seafood and poultry. Vegetarian meal costs were lowest. Our model predicts that the overall effects of price inflation on PPP will vary widely over the next year (Fig.~\ref{fig:price_prediction}C).  While we predicted modest increases for most recipes, with current inflation estimates we expect that prices of about 30\% of recipes will decrease.  

\newpage
\subsection{Recipe optimization }

Combining our compiled nutritional information and predicted recipe prices, we used linear programming to recommend an optimal recipe schedule that PPP might use for the first half of their upcoming academic year distribution (15 weeks). See Table 2.  

The selected plan achieved a total estimated cost of \$25.05 per 4-person family, with a cumulative nutritional content of 512.5 g of protein, 6518.3 mg of calcium, and 385.5 g of fat (mean values of 34.17 g, 434.55 mg, and 25.7 g per serving respectively). As intended, the 15 recipes include at least 3, and no more than 4, recipes from each category.

\begin{table}[ht]
\caption{Results of the 15-week optimized meal plan. }
\label{tab:opt}
    \centering
    \begin{tabular}{|p{3.5cm}|p{3.5cm}|p{3.5cm}|p{3.5cm}|}
        \hline
    \textbf{Beef} & \textbf{Poultry} & \textbf{Seafood} & \textbf{Vegetarian}  \\ \hline
    Beef Fried Rice, \$0.62 & Turkey Chili Dogs, \$1.43 & Salmon Patties, \$0.41 & Bean Chili Cheese Fries, \$0.45  \\ \hline
    Beef Chili Bake, \$2.09 & Egg \& Turkey Sausage Breakfast, \$1.44 & Deconstructed Salmon Roll with Cucumber, \$0.76 & Cauliflower Soup, \$1.30  \\ \hline
    Taco Bowl, \$3.07 & Chicken \& Bean Burritos, \$2.06 & Salmon Croquettes with Mac and Cheese, \$1.60 & Vegetarian Chili Stuffed Baked Potato, \$1.53  \\ \hline
     \quad & Turkey and Cheese Sandwiches \& Soup, \$2.79 & Seafood Chowder, \$2.75 & Macaroni and Cheese, \$2.75  \\ \hline
    \end{tabular}
    \end{table}

\section{Discussion}

The present study developed a pipeline to provide the Power Packs Project, a local nonprofit, with essential information that will help them perform their job more efficiently on a weekly-to-monthly basis. The recommended optimal recipe schedule will also help PPP spend their funds more effectively over the coming year. The pipeline leverages LLMs, operations research (optimization and basic stochastic modeling), and other data science tools. Our intent to work directly with PPP to support, not reinvent, their recipe workflow is a novel aspect of the work which allows it to have immediate real-world impact.  This pipeline can be replicated for work with any local organization fighting food insecurity, further increasing its potential reach.

Beyond the recipe optimization itself, the system makes the PPP team’s work easier by helping them keep track of which recipes and ingredients have been used, and by providing quick access to that information. This improves weekly operational efficiency and supports longer-term planning. The recommended plan reflects a well-balanced distribution across food categories, maintains affordability, and satisfies essential nutritional thresholds. In addition, the pipeline is designed for easy updates: PPP staff can add new recipes to the database over time with minimal effort, ensuring the tool remains useful as their needs evolve.

The string-matching procedure provided an essential means for connecting recipe ingredients with nutritional information, and will continue to offer value as new recipes are added.  While this did require a few rounds of manual intervention, the overall time to obtain these matches was much quicker than if we had trained our own language model \cite{feher2023}.

Some challenges with the pricing data affected our ability to generate accurate price predictions.  Certain wholesale price trends may be influenced by local factors unrelated to national supply and demand. For example, discounts on items nearing expiration at the Central PA Food Bank. Local produce prices may also fluctuate based on seasonality or weather, which, in some cases, could actually benefit PPP if the system is flexible enough to adapt recipe choices accordingly. We observed considerable variation in prices for items such as canned pinto beans and white rice, likely driven by market dynamics or donation-based availability. These fluctuations justified the use of category-specific constraints in the optimization model to balance both affordability and nutritional diversity. Moreover, we tracked invoices from the four main sources of PPP's inventory; others, including seasonal community food drives, may cause PPP's recipe costs to be even lower than we predict.

One limitation of our cost prediction model is its reliance on general inflation estimates, which are inherently uncertain and subject to rapid change. For example, most of our predicted recipe prices rely on the projected price decline in vegetables \cite{inflation-beef/poultry/fruits/seaood/vegetables}; if instead vegetable prices rise in the next several months, our current model will be outdated.  
Fortunately, the model was designed for easy re-use: it can be re-run using updated inflation assumptions as needed. Looking forward, several plausible inflation scenarios could be incorporated into a Monte Carlo framework to provide better estimates of average costs, including confidence intervals.

Several directions for future work remain. We did not employ regression for price prediction due to limited historical depth of invoices; this would become feasible as more data become available.  We could also model the retail costs that families would pay to prepare these meals on their own, optimizing both retail and wholesale costs with a multi-objective approach.  Incorporating data from user meal preference surveys could further enhance our recommended recipe schedule.

Finally, the website infrastructure is being expanded to include dynamically generated nutritional summaries, estimated cost breakdowns, and interactive visualizations of price trends. These features will support transparency, enhance usability, and allow for direct integration with the optimization system. In the longer term, the system could also incorporate data on real-time warehouse management by tracking product quantities, expiration dates, and unit costs.  Such updates would further improve time- and cost-efficiency, helping Power Packs to fulfill their noble mission:  ending hunger among school children over the weekend \cite{ppp-site}. 


\end{document}